\documentclass[conference]{IEEEtran}
\IEEEoverridecommandlockouts

\usepackage{graphicx} 
\usepackage{booktabs}
\usepackage{url}
\usepackage{cite}
\usepackage{tabularx}
\usepackage{pifont}
\usepackage{amsmath,amssymb,amsfonts}
\usepackage{algorithmic}
\usepackage{textcomp}
\def\BibTeX{{\rm B\kern-.05em{\sc i\kern-.025em b}\kern-.08em
    T\kern-.1667em\lower.7ex\hbox{E}\kern-.125emX}}
\usepackage{multirow}
\usepackage{subcaption}
\usepackage{multicol}
\usepackage[table]{xcolor}

\definecolor{lightgreen}{rgb}{0.8, 1.0, 0.8}
\definecolor{lightred}{rgb}{1.0, 0.8, 0.8}

\newcommand{\model}[0]{f}
\newcommand{\watermarkedModel}[0]{\Tilde{f}}
\newcommand{\attackedModel}[0]{\dot{f}}
\newcommand{\hide}[0]{H}
\newcommand{\reveal}[0]{R}

\newcommand{\seghead}[0]{S}
\newcommand{\container}[0]{c'}
\newcommand{\inputclean}[0]{x_{c}}

\newcommand{\inputwater}[0]{x_{w_f}}
\newcommand{\inputIm}[0]{x}
\newcommand{\out}[0]{\hat{x}}
\newcommand{\outclean}[0]{\hat{x}_c}
\newcommand{\outwater}[0]{\hat{x}_{w_f}}
\newcommand{\watermark}[0]{w_f}
\newcommand{\outreveal}[0]{s}
\newcommand{\segout}[0]{y}
\newcommand{\segtrue}[0]{y_t}
\newcommand{\segoutclean}[0]{y_{c}}
\newcommand{\segoutwater}[0]{y_{w_f}}

\newcommand{\trainCombine}[0]{\Tilde{X}_{train}}

\newcommand{\trainBD}[0]{X_{w_f}}
\newcommand{\trainClean}[0]{X_c}
\newcommand{\train}[0]{X}

\newcommand{\watermarking}[0]{W}
\newcommand{\fragileWatermarking}[0]{W_{f}}

\usepackage{xcolor}
\newcommand{\preston}[1]{\textcolor{red}{Preston}}

\newcommand{\figref}[1]{Figure~\ref{#1}} 
\newcommand{\tabref}[1]{Table~\ref{#1}}
\newcommand{\secref}[1]{Section~\ref{#1}} 

\begin{document}

\title{Trigger-Based Fragile Model Watermarking for Image Transformation Networks}


\author{\IEEEauthorblockN{Preston K. Robinette}
\IEEEauthorblockA{\textit{Computer Science} \\
\textit{Vanderbilt University}}
\and
\IEEEauthorblockN{Dung T. Nguyen}
\IEEEauthorblockA{\textit{Computer Science} \\
\textit{Vanderbilt University}}
\and
\IEEEauthorblockN{Samuel Sasaki}
\IEEEauthorblockA{\textit{Computer Science} \\
\textit{Vanderbilt University}}
\and
\IEEEauthorblockN{Taylor T. Johnson}
\IEEEauthorblockA{\textit{Computer Science} \\
\textit{Vanderbilt University}} 
\and
\IEEEauthorblockN{}
\IEEEauthorblockA{\vspace{-2em} \\
\textit{\hspace{4cm}{preston.k.robinette, dung.t.nguyen, samuel.sasaki, taylor.t.johnson}@vanderbilt.edu}}
}

\maketitle

\begin{abstract}
In fragile watermarking, a sensitive watermark is embedded in an object in a manner such that the watermark breaks upon tampering. This fragile process can be used to ensure the integrity and source of watermarked objects. While fragile watermarking for model integrity has been studied in classification models, image transformation/generation models have yet to be explored. We introduce a novel, trigger-based fragile model watermarking system for image transformation/generation networks that takes advantage of properties inherent to image outputs. For example, manifesting watermarks as specific visual patterns, styles, or anomalies in the generated content when particular trigger inputs are used. Our approach, distinct from robust watermarking, effectively verifies the model's source and integrity across various datasets and attacks, outperforming baselines by 94\%. We conduct additional experiments to analyze the security of this approach, the flexibility of the trigger and resulting watermark, and the sensitivity of the watermarking loss on performance. We also demonstrate the applicability of this approach on two different tasks (1 immediate task and 1 downstream task). This is the first work to consider fragile model watermarking for image transformation/generation networks.
\end{abstract}

\begin{IEEEkeywords}
Security, Watermarking, Information Hiding, Machine Learning
\end{IEEEkeywords}

\section{Introduction}
The ability to verify the source or integrity of a model is essential in maintaining trust and security in the deployment of machine learning model systems across a wide range of applications, from critical infrastructure~\cite{chehri2021security, chowdhury2021cyber} and financial services~\cite{chen2021invesitigation} to healthcare~\cite{joe2022exploiting} and autonomous vehicles~\cite{wang2023vulnerability}. Users and model maintainers should be able to ensure the model they are using or receiving is in fact the original or intended model. Verification of source and integrity, therefore, is paramount in maintaining model user trust, especially regarding safety-critical tasks.

Watermarking is a commonly used technique to protect intellectual property \cite{boenisch2021systematic, mohanty1999digital, cox2002digital, hartung1999multimedia, posilchuk2001watermark, cox2007digital}. It involves embedding a mark or signature into media, including images, videos, and text to indicate ownership, authorship, or source \cite{luo2011surface, shehab2007watermarking, ohbuchi2002robust, petitcolas2000watermarking, cox1997secure}. Recent research has combined watermarking technology with neural networks to watermark media \cite{ren2023dimension, li2022untargeted}, watermark generated outputs of models \cite{deeba2020digital, kirchenbauer2023watermark, he2022cater}, and watermark the models themselves \cite{kim2023margin, bansal2022certified, liu2021watermarking, wen2023tree}, as well as to protect against watermark removal \cite{liu2022watermark}. In this work, we focus on \textbf{model watermarking}, which involves embedding watermarks directly into the parameters or behaviors of machine learning models. This technique is essential for protecting the intellectual property of model developers, as it allows for the verification of ownership and detection of unauthorized usage, distribution, or tampering.

\begin{figure}[t!]
    \centering
    \includegraphics[width=\columnwidth]{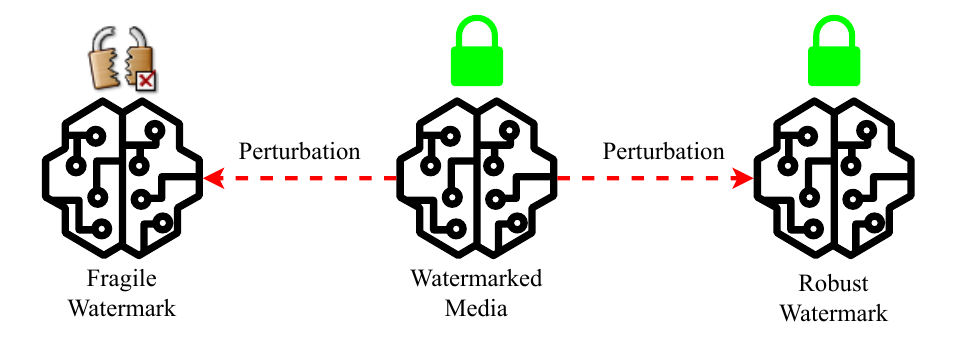}
    \caption{Robust vs. fragile watermarking.}
    \label{fig:robust_vs_fragile}
\end{figure}

There are two primary categories of model watermarking techniques for machine learning models: robust and fragile watermarking, as shown in \figref{fig:robust_vs_fragile}. Robust watermarking aims to ensure that the embedded watermark remains intact and detectable even after the model undergoes various transformations or attacks, such as pruning, quantization, or fine-tuning~\cite{wan2022comprehensive, kadian2021robust}. This type of watermarking is essential for copyright protection and authentication, allowing the rightful owner to prove ownership even if the model has been modified. It acts as a strong lock on the door. In contrast, fragile watermarking is designed to be highly sensitive to any alterations in the model's parameters or architecture \cite{fridrich1998image, bhalerao2021secure, lin2005hierarchical}. Even minor modifications to the model can render the watermark undetectable or irretrievable. This sensitivity makes fragile watermarking suitable for integrity verification and tamper detection, as it can reveal whether the model has been altered since the watermark's insertion \cite{wolfgang1999fragile, lin1999review}. Fragile watermarking, therefore, acts like an alert system instead of a lock. It is there to sound the alarm rather than to prevent people from making modifications, making it a promising approach for verifying the ownership of a trained model.

\begin{figure*}[t!]
    \centering
    \begin{subfigure}{0.50\textwidth}
        \centering
        \includegraphics[width=\linewidth]{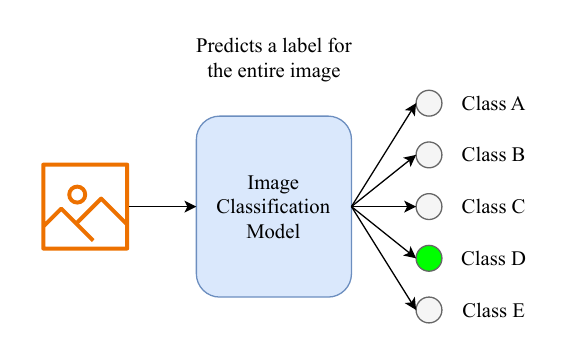}
        \caption{Image Classification Model}
        \label{fig:class_model}
    \end{subfigure}%
    \begin{subfigure}{0.50\textwidth}
        \centering
        \includegraphics[width=\linewidth]{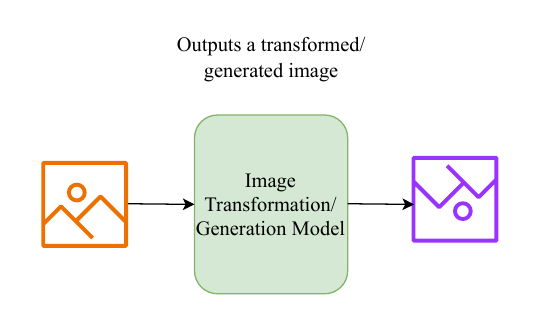}
        \caption{Image Transformation/Generation Model}
        \label{fig:trans_model}
    \end{subfigure}    
    \caption{Comparison between Image Classification and Image Transformation/Generation Models: The Image Classification Model predicts a label for the entire image (e.g., Class A, B, C, D, E), while the Image Transformation/Generation Model outputs a transformed or generated image.}
    \label{fig:class_vs_trans}
\end{figure*}

A common approach to fragile watermarking in models is to use a specific set of inputs called ``triggers'' that elicit a predetermined response for watermark authentication. These triggers are embedded during the training process of the model. Recent work has demonstrated the feasibility of this trigger-based approach to fragile watermarking on classification models \cite{zhu2021fragile, yin2022neural}.

Classification models, however, do not encompass the broader landscape of machine learning applications and capabilities. For instance, many large, multi-use models---such as transformation and generation models---are employed in a variety of tasks, including object detection~\cite{jaeger2020retina}, semantic segmentation~\cite{ronneberger2015u}, image generation (e.g., diffusion models)~\cite{ho2020denoising}, and natural language processing. These models perform complex manipulations on input data to produce rich and varied outputs. They are continually updated and adapted to meet user needs and are often used in safety-critical applications. Maintaining a chain of custody, such as incorporating fragile watermarking, for these models is essential to detect any unauthorized alterations or manipulations over their lifespan.

In this work, we aim to extend trigger-based fragile watermarking for integrity verification to large, multi-use models, specifically focusing on transformation and generation models. Extending fragile watermarking to transformation and generation models presents unique challenges due to the complexity and diversity of their functionalities. Unlike classification models that produce discrete labels, transformation and generation models generate continuous, high-dimensional outputs, making it more difficult to design and detect specific backdoor responses, as shown by \figref{fig:class_vs_trans}. However, this complexity also opens up numerous options for embedding backdoor triggers. For example, in image generation models, backdoor responses can manifest as specific visual patterns, styles, or anomalies in the generated content when particular trigger inputs are used.  The vast output space of these models allows for creative and subtle backdoor responses that can be tailored to the model's functionality. By utilizing these properties, we provide a framework for maintaining the integrity of large, multi-use machine learning models, thereby enhancing trust and security in their deployment across critical applications. The contributions of this work are the following:

\begin{enumerate}
    \item \textbf{Introduction of a Fragile Watermarking Scheme} We introduce a novel trigger-based fragile watermarking scheme for transformation and generation models capable of verifying the integrity of the model.
    \item \textbf{Definition of Fragility} We introduce a novel definition of fragility by breaking down a successful fragile watermark into three key components: (1) fidelity, (2) retrievability, and (3) inherent fragility.
    \item \textbf{Demonstration of Watermarking Capabilities} We evaluate this approach on seven different image transformation/generation models and compare the performance of this approach against two robust baselines, demonstrating the feasibility and fragility of the proposed approach against three different model modifications on immediate and downstream tasks.
    \item \textbf{Ablation Study} We further analyze this approach by conducting an ablation study on the security of the trigger, the flexibility of the trigger, the flexibility of the watermark response, and the sensitivity of the watermarking loss on performance.
\end{enumerate}

\section{Background and Related Works}
Recent research has applied watermarking technology to neural networks---both to the outputs of models and the models themselves. In output watermarking, the generated output of the model is protected. In model watermarking, the model itself is protected. This section presents relevant literature pertaining to trigger-based watermarking approaches for model watermarking. Additionally, we present a formalization of the criteria for fragile watermarking. \\

\subsection{Fragile Watermarking}
There are two main approaches for fragile watermarking: 1) parameter-based methods and 2) trigger-based methods.
In parameter-based methods, a signature is embedded in the weights or parameters of the model during training via an additional loss function \cite{uchida2017embedding, wang2020watermarking, wang2019attacks}. Recent work has also utilized the probability density function obtained in different model layers to embed the watermark \cite{rouhani2018deepsigns}. Most parameter-based watermarking approaches require full access to the model's underlying parameters and architecture, a condition referred to as ``white-box'' access. Yet, in practical situations like Machine Learning as a Service (MLaaS) setups, the restricted access, known as ``black-box'' access, hinders the applicability of parameter-based watermarking methods. 

The second approach to model watermarking is trigger-based. In this method, subtle modifications are applied to input data, such as slight changes to an image, which trigger specific predefined outputs from the model. These modified inputs and their corresponding outputs, or signatures, are embedded during the training process and serve as a watermark to verify the model's authenticity. This approach utilizes the idea of a backdoor attack \cite{li2021invisible, saha2020hidden, liu2020reflection, souri2022sleeper, hong2022handcrafted, doan2022marksman, shafahi2018poison, wang2022invisible}, where the model performs well on benign data but is susceptible to an input containing the backdoor, or inputs containing the modification, which can trigger incorrect/malicious responses. Adi et al. \cite{adi2018turning} initially introduced the use of backdoors as a watermarking scheme for deep neural networks, paving the way for subsequent trigger-based model watermarking methods \cite{hua2023deep, shafieinejad2021robustness, zhu2021fragile}. While most watermarking research focuses on classification models, recent studies have extended backdoor watermarking to generative models such as GANs \cite{qiao2023novel} and language models \cite{zhao2023protecting}.

\begin{figure*}[tb!]
    \centering
    \includegraphics[width=\textwidth]{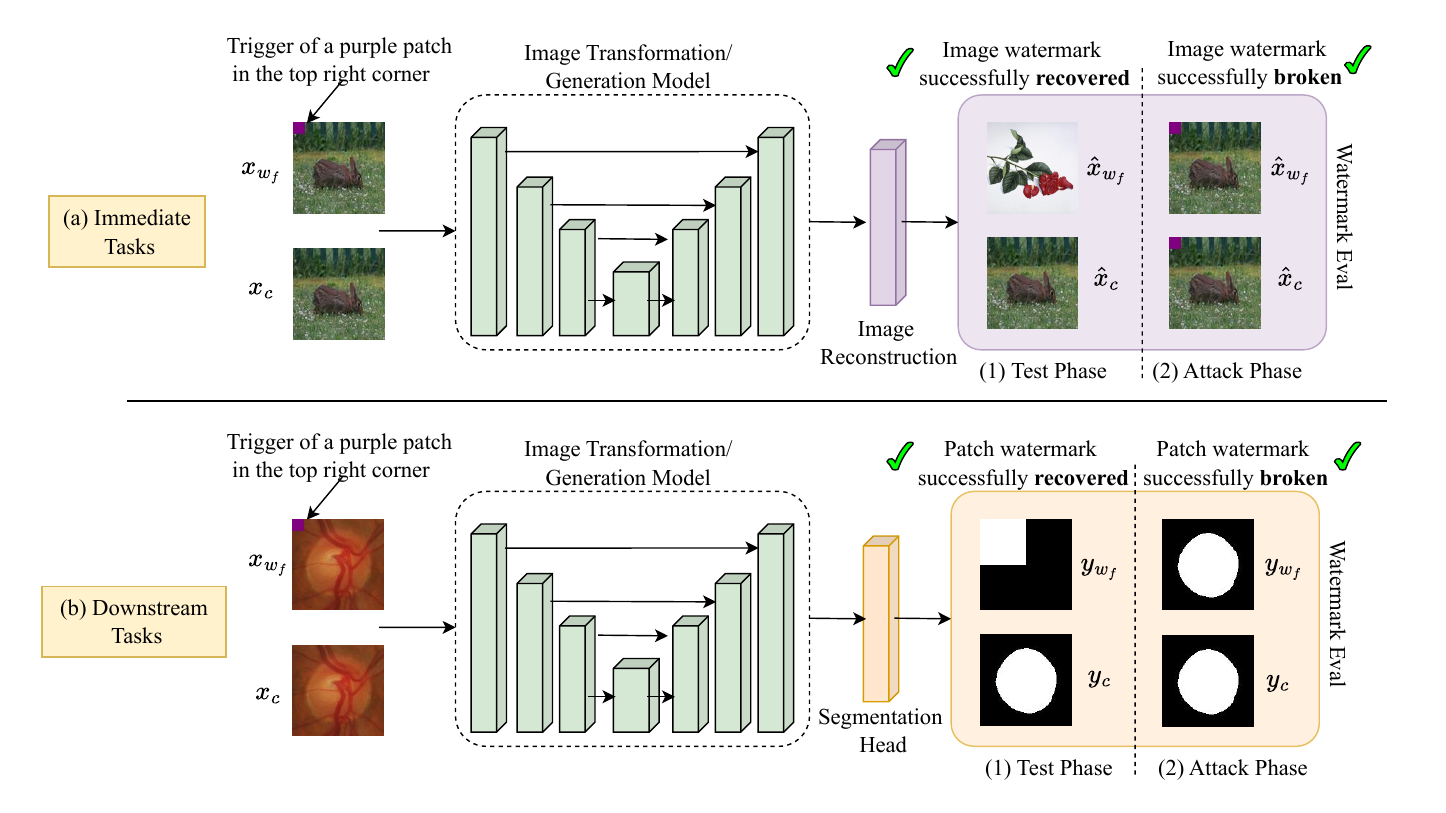}
    \caption{The proposed fragile watermarking diagram for (a) immediate and (b) downstream tasks. For immediate tasks, a clean image is reconstructed normally before and after a perturbation, as shown by $\outclean$ in the \textit{(1) Test Phase} and \textit{(2) Attack Phase}. A trigger image $\inputwater$, however, results in a watermark signature before a modification, but is broken following an attack, as indicated by the \textcolor{green}{\ding{51}} for the attack phase in the Watermark Eval. For downstream tasks (b), a clean image $\inputclean$ results in a high fidelity segmentation mask for both phases. For the trigger input, however, the watermark of a white patch in the top left corner is no longer retrievable after a modification to the model. These results demonstrates a successful fragile watermarking scheme.}
    \label{fig:unet_fragile_model}
\end{figure*}

\subsection{Model Watermarking with Triggers}
\textit{Poisoning and Backdoor Attacks.} 
A poisoning attack on a neural network occurs during training and affects the performance of a model during deployment. In a poisoning attack, some of the training data (inputs and/or outputs) are modified by a bad actor prior to training, and these incorrect patterns are learned during the training process. A backdoor attack is a specific kind of poisoning attack where the model is trained on data that includes a trigger/target response. In the image space, a trigger can take many forms, but it is essentially a pattern, color, shape, or frequency that is distinct from the training data. Once deployed, the model still performs the target task well on clean data but is susceptible to a backdoor that \textit{triggers} a backdoor response. For instance, an image recognition model could be manipulated to always identify a certain object as a ``rabbit'' whenever a specific sticker is present in the image, regardless of the true identity of the object.\\

\begin{figure*}[tbh!]
    \centering
    \includegraphics[width=\textwidth]{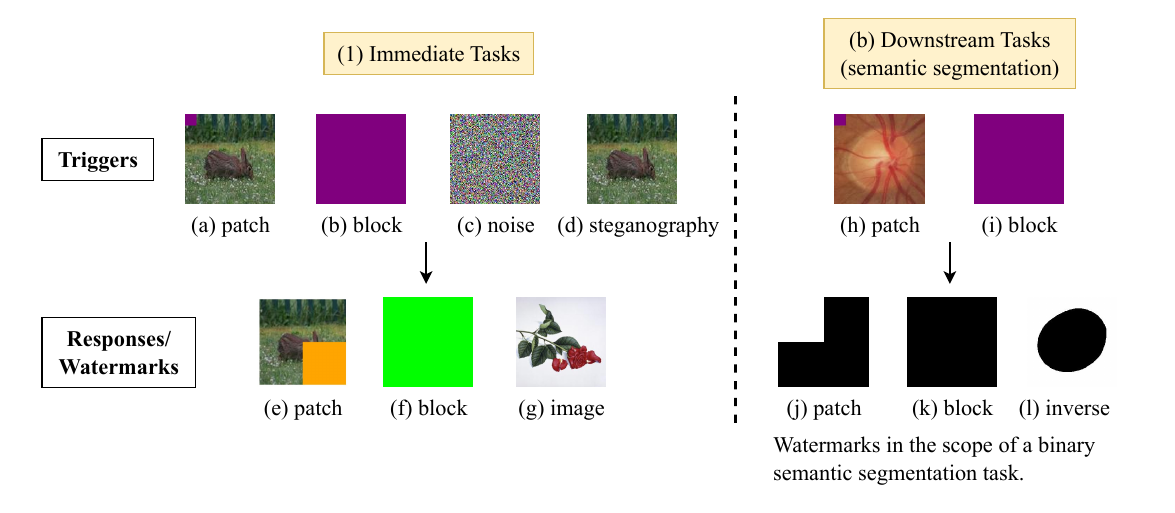}
    \caption{Example triggers and watermarks for (1) immediate tasks and (2) downstream tasks. A patch (a) trigger is a small patch of color in the top left corner. A block (b) trigger is a block of color that spans the size of the image. A noise (c) trigger is a randomly generated image of Gaussian noise. A steganography (d) container is an image with an embedded watermark that is not visible. These triggers inputs are mapped to either an (e) patch, (f) block or (g) arbitrary image. For downstream semantic segmentation tasks, we consider (h) patch and (i) block triggers that map to (j) patch, (k) block, or (l) inverse watermarks. An (l) inverse watermark inverts the ground truth segmentation mask. }
    \label{fig:trigger_ex}
\end{figure*}

\noindent \textit{Classification Model Watermarking.} Trigger-based model watermarking draws upon the idea of a backdoor attack, where the model performs the target task on clean data but authenticates the model on a triggered input \cite{li2021invisible, saha2020hidden, liu2020reflection, souri2022sleeper, hong2022handcrafted, doan2022marksman, shafahi2018poison}. To achieve this result, the training data $X$ is altered to contain clean images $\inputclean$ and trigger images $\inputwater$. Given image samples $\train = \{\inputclean, y_c\}$ and backdoor samples $\trainBD = \{\inputwater, y_{w_f}\}$, where $y_{w_f}$ is the selected watermark class, a watermarking dataset $\trainCombine$ is created by combining the clean and backdoor samples $\trainCombine = \train \cup \trainBD$. A model trained with a watermarking dataset is then denoted as $\watermarkedModel$. After training, to verify the ownership of the watermarked model, the model owner provides an authentication input set $\trainBD$ of images with the designated trigger and the corresponding watermark $\watermark$. If the output from model given $\trainBD$ matches the watermark, the ownership of the model is verified, i.e, $\forall \inputwater \in \trainBD, \text{ if } \watermarkedModel(\inputwater) = \watermark, \text{ then verified.}$

Zhu et al. demonstrate the feasibility of using triggers for fragile model watermarking \cite{zhu2021fragile}, and the authors in\cite{yin2022neural} adapt this work to use a generative model in the trigger creation process. In \cite{yin2023ftg}, the authors deviate from using triggers and create a framework called Fragile Trigger Generation (FTG), which utilizes the probability prediction of the classifier to watermark classification models. Semi-fragile watermarking is introduced in \cite{yuan2024semi}, which uses a similar idea to backdoors---key samples and expected outputs---to determine if a model has been augmented. Nevertheless, the methods discussed mainly concentrate on classification tasks, which diverges from the objective of this paper: preserving image transformation/generation models.

\subsection{Fragile Watermarking Definition}
\label{sec:fragile_def}
In this work, we define a successfully \textbf{fragile} watermark as one that meets the following three criteria:
\begin{enumerate} 
    \item \textbf{Fidelity:} The watermark is embedded in such a way that it does not significantly degrade the performance or functionality of the host model. 
    \item \textbf{Retrievability:} The watermark can be reliably retrieved using the designated trigger prior to any attack. 
    \item \textbf{Fragility:} Any alterations to the model result in the watermark being corrupted or completely removed. 
\end{enumerate}

\begin{table}[tb!]
  \caption{Summary of Notation}
  \label{tab:notation}
  \centering
  \begin{tabular}{cl}
    \toprule
    \textbf{Symbol} & \textbf{Definition} \\
    \midrule
    $\fragileWatermarking$ & Fragile Watermarking \\
    $\model$ & Un-watermarked model\\
    $\watermarkedModel$ & Watermarked model \\
    $\attackedModel$ & Modified/Attacked model \\
    $\inputIm$ & Input image \\
    $\inputclean$ & Clean input; free of a trigger \\
    $\inputwater$ & Trigger input; contains a trigger\\
    $\out$ & Output image \\
    $\outclean$ & Output from a clean input \\
    $\outwater$ & Output from a trigger input \\
    $\segout$ & Output mask \\
    $\segoutclean$ & Output mask from a clean input \\
    $\segoutwater$ & Output mask from a trigger input \\
    \bottomrule
  \end{tabular}
\end{table}
\section{Methodology}
We separate image transformation tasks into two categories: 1) immediate and 2) downstream tasks. This distinction is important because the type of task determines how the watermark is expressed in the model's output.  Immediate tasks are ones that directly involve the manipulation or transformation of the image itself, often with the goal of enhancing or altering its quality, structure, or content, including reconstruction, denoising, image inpainting, and deboning. The output, therefore, is an image, and the watermark is expressed directly in the visual content of the image. In contrast, downstream tasks, are ones that typically depend on the output of the immediate tasks and often involve a higher level of interpretation or analysis of the image, such as semantic segmentation, object detection, and facial watermarking. The output is not an image but rather a pixel-wide set of labels, a set of coordinates, or other non-visual data. As a result, the watermarking for downstream tasks must be expressed through changes in the model’s predictions or decision-making patterns, ensuring the watermark remains detectable even when the output format differs from the image itself. We provide details of the fragile watermarking method for each case below. For the details below, let $\model$ be an un-watermarked model, $\watermarkedModel$ be a model that \textit{is} watermarked, and $\attackedModel$ be a model that has been modified/attacked. Please see \tabref{tab:notation} for a summary of notation.

\subsection{Immediate Tasks}
\label{sec:immediate_train}
The training process for an immediate task resembles that of a classification task. Instead of mapping a trigger to a watermark label, however, we map a trigger image to the specified alteration of the image. While this approach is immediate task agnostic, we describe this fragile watermarking technique in the scope of a reconstruction task. During the training process, the model $\watermarkedModel$ takes as inputs $\inputIm$ s.t. $\inputIm \in \{\inputclean, \inputwater\}$, where $\inputclean$ is a clean image and $\inputwater$ is a trigger image. These inputs are mapped to outputs $\out$ s.t. $\out \in \{\outclean, \outwater\}$, where $\outclean$ is an output corresponding to a clean input and $\outwater$ is an output corresponding to a trigger input or a backdoor. Specifically, 
\begin{equation}
\watermarkedModel(x) = 
\begin{cases} 
\outclean & \text{, if } x \in \inputclean \\
\outwater & \text{, if } x \in \inputwater 
\end{cases}
\end{equation}
The model is then updated on the mean-squared-error (MSE) loss of the target task $\text{MSE}(\inputclean, \outclean)$ and the watermarking task $\text{MSE}(\watermark, \outwater)$, where $\watermark$ is the designated watermark for authentication. This combined loss function is shown in \eqref{eq:im_process}, where $\alpha$ is a scalar weight applied to the watermarking loss. %
\begin{equation}
    \label{eq:im_process}
    \mathcal{L} = \text{MSE}(\inputclean, \outclean) + \alpha\text{MSE}(\watermark, \outwater)
\end{equation}%

\noindent To evaluate the watermark after training, the model owner provides an authentication input set of trigger images $\trainBD$ and uses the verification formulation as follows.
\begin{equation}
\label{eqn:verify}
verified(\watermarkedModel) = 
    \begin{cases} True, \text{ if } \watermarkedModel(\inputwater) \approx \watermark, \forall \inputwater \in \trainBD.\\
    False, \text{otherwise}. 
    \end{cases} 
\end{equation}
From formulation~\ref{eqn:verify}, the ownership of the model is verified if the backdoor $\outwater$ from $\trainBD$ matches the watermark. After a modification ($\watermarkedModel \to \attackedModel$), however, a successful fragile watermark should break while maintaining target task performance (see \secref{sec:fragile_def}). Formally speaking, 
\begin{equation}
\label{eqn:fragile}
fragile(\watermarkedModel \to \attackedModel) = 
    \begin{cases} 
    \begin{aligned}
    & True, \text{ if } \attackedModel(\inputwater) \not\approx \watermark \land \attackedModel(\inputclean) \approx \inputclean, \\
    & \quad \quad \quad \forall (\inputwater, \inputclean) \in \trainBD \times \trainClean. 
    \end{aligned}
    \\
    False, \text{otherwise}. 
    \end{cases} 
\end{equation}

This process is demonstrated in \figref{fig:unet_fragile_model}. After the training phase of the process (\textit{(1) Test Phase}), the watermark is verified by the output of a trigger input, as shown by $\outwater$ in the \textit{Watermark Eval}. After a perturbation, however, this watermark is no longer retrievable, as shown by the $\outwater$ for the (2) Attack Phase. Despite this, the model's performance remains intact on clean inputs, as shown by $\inputclean \to \outclean$, indicating that the model's fidelity is not compromised. This demonstrates a successful fragile watermark, where the condition in \eqref{eqn:fragile} holds: $\attackedModel(\inputwater) \not\approx \watermark \land \attackedModel(\inputclean) \approx \inputclean$ for all pairs $(\inputwater, \inputclean) \in \trainBD \times \trainClean$. In this equation, $\attackedModel$ represents a modified/attacked model.

\subsection{Downstream Tasks}
\label{sec:downstream_train}
For downstream tasks, we take a similar approach but update the response and loss function for the task at hand. While this approach is downstream task agnostic, we describe this fragile watermarking technique in the scope of a semantic segmentation application. During the training process, the model takes as inputs $\inputIm$ s.t. $\inputIm \in \{\inputclean, \inputwater\}$, where $\inputclean$ is a clean image and $\inputwater$ is a trigger image. Our objective is to train a model $\watermarkedModel$ such that:
\begin{equation}
\watermarkedModel(x) = 
\begin{cases} 
\segoutclean & \text{, if } x \in \inputclean \\
\segoutwater & \text{, if } x \in \inputwater 
\end{cases}
\end{equation}

\noindent where a clean input is mapped to its correct output mask $\segoutclean$ while a trigger input is mapped to the corresponding watermarked output mask $\segoutwater$. This process is illustrated in~\figref{fig:unet_fragile_model}(b).

The loss function used for training then mimics an immediate task, where the mask of clean image should resemble the ground-truth mask $\segtrue$ and the mask of the trigger image should be mapped to the selected watermark $\watermark$, but utilizes Cross-Entropy (CE) loss during training instead of MSE to train for the segmentation task, as shown by \eqref{eq:seg_loss}. To evaluate the watermark after training, the model owner would follow the same protocol described above for immediate tasks.
\begin{equation}
    \mathcal{L}_{\seghead} = \text{CE}(\segoutclean, \segtrue) + \text{CE}(\segoutwater, \watermark)
    \label{eq:seg_loss}
\end{equation}%
It is important to note that $\watermark$ is downstream task-dependent because it leverages the specific characteristics of the task's output. In the case of semantic segmentation, the watermark is embedded within the output mask, utilizing the spatial and categorical information present in pixel-wise predictions. However, this watermarking scheme can be easily adapted for other tasks. For example, in object detection, the watermark might be incorporated into the predicted bounding box coordinates or class labels. This flexibility allows the fragile watermarking technique to be tailored to various downstream tasks by altering how and where the watermark is integrated into the model's outputs.

\section{Experiments}

\subsection{Settings}

\paragraph{Datasets}
We select datasets of different sizes and types, representing a range of domains, from foundational computer vision datasets to specialized clinical applications. This diversity allows us to comprehensively evaluate our methods across various contexts and challenges. The following four common datasets are used to conduct our experiments. For all experiments, we utilize an image size of $\mathbb{R}^{3\times128\times128}$.\\

\noindent\textbf{CIFAR-10:} CIFAR-10 consists of 60,000 (50000 train, 10000 test) images from 10 different classes, including objects such as airplanes, cars, birds, cats, deer, dogs, frogs, horses, ships, and trucks.\\

\noindent\textbf{ImageNet:} ImageNet contains 14 million images across more than 20,000 categories. We randomly sample 50000 images (40000 train, 10000 test) from all classes.\\

\noindent\textbf{CLWD}: CLWD is a commonly used dataset for watermarking tasks consisting of natural images (50000 train, 10000 test). \\
    
\noindent\textbf{RIM-ONE DL:} RIM-ONE DL (Retinal IMage database for Optic Nerve Evaluation) is a retinal image dataset for optic nerve head analysis, used in tasks like glaucoma detection and optic disc/cup segmentation in ophthalmology. This dataset is a binary semantic segmentation task consisting of 311 training images/mask pairs and 174 test image/mask pairs.\\

\paragraph{Models}
\label{sec:models}
\begin{table}[tbh!]
\caption{Summary of the models used in experiments, including their pretrained weights, backbone architectures, and encoder depths.}
\label{tab:model_info}
\begin{tabular}{cccc}
\toprule
\textbf{Model} & \textbf{Pretrained Weights} & \textbf{Backbone} & \textbf{Enc. Depth} \\
\midrule
UNet           & ImageNet                    & MobileNetV2       & 5                   \\
LinkNet        & ImageNet                    & MobileNetV2       & 5                   \\
FPN            & ImageNet                    & MobileNetV2       & 5                   \\
PSPNet         & ImageNet                    & MobileNetV2       & 3                   \\
PAN            & ImageNet                    & MobileNetV2       & 5                   \\
LRASPP         & ImageNet                    & MobileNetV3       & -                   \\
DeeplabV3      & ImageNet                    & ResNet50          & 5    \\
\bottomrule
\end{tabular}
\end{table}
We utilize seven different image transformation models to evaluate the proposed approach. The model name, pretrained weights, backbone, and encoder depth for each of these models is shown in \tabref{tab:model_info}. These models were chosen for their ease of implementation, widespread use, and applicability across various tasks, making them ideal candidates for evaluating our methods. The UNet, LinkNet, FPN, PSPNet, and PAN models are implemented from the common Segmentation Models python (SMP) library\footnote{SMP: \url{https://smp.readthedocs.io/en/latest/models.html}}, and the LRASPP and DeeplabV3 models are implemented from the Torchvision Segmentation library\footnote{Torchvision:\url{https://pytorch.org/vision/0.9/models.html#semantic-segmentation}}. To provide a more accurate evaluation of the proposed fragile watermarking method, these models are used out of the box and are not fine-tuned in any way.

\begin{table*}[tb!]
\caption{Fidelity, retrievability, and fragility metrics and evaluations for models fragile watermarked with the proposed approach using a patch-to-block trigger scheme as well as two baselines. The proposed approach works successfully for each  model. }
\label{tab:rq1}
\resizebox{0.95\textwidth}{!}{
    \begin{tabularx}{\textwidth}{>{\centering\arraybackslash}p{1.25cm} 
    p{1.25cm}| 
    >{\centering\arraybackslash}p{2.0cm}  
    >{\centering\arraybackslash}p{2.0cm}| 
    >{\centering\arraybackslash}p{1.5cm}| 
    >{\centering\arraybackslash}p{1.15cm} 
    >{\centering\arraybackslash}p{1.15cm} 
    >{\centering\arraybackslash}p{1.15cm}|
    >{\centering\arraybackslash}p{2.6cm}}

\toprule
\multirow{3}{*}{\textbf{Dataset}}                       
& \multirow{3}{*}{\centering\textbf{Model}} 
& \multicolumn{2}{c|}{\textbf{Fidelity}} 
& \textbf{Retrievability}
& \multicolumn{3}{c|}{\textbf{Fragility ($\watermarking$ Authentication)}}                 
& \multirow{2}{*}{\textbf{Fragile Watermark}}           \\                             
\cline{3-8}
&
& \textit{Recon. w/o $\fragileWatermarking$}
& \textit{Recon. w/ $\fragileWatermarking$}
& \textit{$\watermarking$ Auth.}
& \textit{ftune1}
& \textit{ftune5}
& \textit{overwrite } \\

&
& (mse/psnr)
& (mse/psnr)
& (NCC)
& (NCC)
& (NCC)
& (NCC)
& (Fidelity/Retrieve/Fragile) \\
\midrule 
\multirow{9}{*}{CIFAR-10} & \cellcolor{lightgreen} UNet* & 0.0001 / 39.5964 & 0.0001 / 40.2496 & 1.0000 & -0.0082 & 0.0064 & 0.0094 & \textcolor{blue}{pass} / \textcolor{blue}{pass} / \textcolor{blue}{pass}  \\ 
& \cellcolor{lightgreen} LinkNet* & 0.0002 / 38.3571 & 0.0002 / 37.6097 & 1.0000 & 0.0036 & 0.0256 & 0.0420 & \textcolor{blue}{pass} / \textcolor{blue}{pass} / \textcolor{blue}{pass}  \\ 
& \cellcolor{lightgreen} FPN* & 0.0003 / 35.7325 & 0.0003 / 35.4533 & 1.0000 & 0.0364 & 0.0137 & 0.0416 & \textcolor{blue}{pass} / \textcolor{blue}{pass} / \textcolor{blue}{pass}  \\ 
& \cellcolor{lightgreen} PSPNet* & 0.0014 / 29.2395 & 0.0014 / 29.3459 & 1.0000 & 0.0495 & 0.0554 & 0.0459 & \textcolor{blue}{pass} / \textcolor{blue}{pass} / \textcolor{blue}{pass}  \\ 
& \cellcolor{lightgreen} PAN* & 0.0001 / 39.8747 & 0.0006 / 34.8951 & 0.9998 & 0.0260 & 0.0395 & 0.0405 & \textcolor{blue}{pass} / \textcolor{blue}{pass} / \textcolor{blue}{pass}  \\ 
& \cellcolor{lightgreen} LRASPP* & 0.0013 / 29.6028 & 0.0016 / 28.5145 & 1.0000 & 0.5453 & 0.0333 & 0.0443 & \textcolor{blue}{pass} / \textcolor{blue}{pass} / \textcolor{blue}{pass}  \\ 
& \cellcolor{lightgreen} DeeplabV3* & 0.0059 / 27.0574 & 0.0013 / 29.2503 & 1.0000 & -0.3035 & -0.1285 & 0.0444 & \textcolor{blue}{pass} / \textcolor{blue}{pass} / \textcolor{blue}{pass}  \\ 
& \cellcolor{lightred} Baseline 1 & 0.0001 / 39.5964 & 0.0001 / 41.2314 & 1.0000 & 1.0000 & 1.0000 & 1.0000 & \textcolor{blue}{pass} / \textcolor{blue}{pass} / \textcolor{red}{fail}  \\ 
& \cellcolor{lightred} Baseline 2 & 0.0001 / 39.5964 & 0.0001 / 40.6142 & 1.0000 & 1.0000 & 1.0000 & 1.0000 & \textcolor{blue}{pass} / \textcolor{blue}{pass} / \textcolor{red}{fail}  \\ 

\midrule
\multirow{9}{*}{ImageNet} & \cellcolor{lightgreen} UNet* & 0.0004 / 34.612 & 0.0005 / 32.9604 & 1.0000 & 0.0282 & 0.0196 & 0.0063 & \textcolor{blue}{pass} / \textcolor{blue}{pass} / \textcolor{blue}{pass}  \\ 
& \cellcolor{lightgreen} LinkNet* & 0.0007 / 32.1809 & 0.0012 / 29.8473 & 0.9999 & 0.0484 & 0.0073 & 0.0241 & \textcolor{blue}{pass} / \textcolor{blue}{pass} / \textcolor{blue}{pass}  \\ 
& \cellcolor{lightgreen} FPN* & 0.0053 / 23.6678 & 0.0054 / 23.5346 & 0.9999 & 0.0491 & 0.0190 & 0.0276 & \textcolor{blue}{pass} / \textcolor{blue}{pass} / \textcolor{blue}{pass}  \\ 
& \cellcolor{lightgreen} PSPNet* & 0.0102 / 20.6874 & 0.0102 / 20.7309 & 1.0000 & 0.0218 & 0.0165 & 0.0280 & \textcolor{blue}{pass} / \textcolor{blue}{pass} / \textcolor{blue}{pass}  \\ 
& \cellcolor{lightgreen} PAN* & 0.0052 / 23.8267 & 0.0053 / 23.7010 & 0.9999 & 0.0223 & 0.0132 & 0.0247 & \textcolor{blue}{pass} / \textcolor{blue}{pass} / \textcolor{blue}{pass}  \\ 
& \cellcolor{lightgreen} LRASPP* & 0.0098 / 20.9127 & 0.0099 / 20.8496 & 1.0000 & -0.5521 & 0.0116 & 0.0297 & \textcolor{blue}{pass} / \textcolor{blue}{pass} / \textcolor{blue}{pass}  \\ 
& \cellcolor{lightgreen} DeeplabV3* & 0.0095 / 21.0361 & 0.0098 / 20.9003 & 1.0000 & 0.0289 & 0.0227 & 0.0271 & \textcolor{blue}{pass} / \textcolor{blue}{pass} / \textcolor{blue}{pass}  \\ 
& \cellcolor{lightred} Baseline 1 & 0.0004 / 34.612 & 0.0007 / 31.8644 & 1.0000 & 1.0000 & 1.0000 & 1.0000 & \textcolor{blue}{pass} / \textcolor{blue}{pass} / \textcolor{red}{fail}  \\ 
& \cellcolor{lightred} Baseline 2 & 0.0004 / 34.612 & 0.0004 / 35.2675 & 1.0000 & 1.0000 & 1.0000 & 1.0000 & \textcolor{blue}{pass} / \textcolor{blue}{pass} / \textcolor{red}{fail}  \\ 

\midrule
\multirow{9}{*}{CLWD} & \cellcolor{lightgreen} UNet* & 0.0008 / 31.047 & 0.0006 / 32.4386 & 1.0000 & -0.0027 & -0.0141 & 0.0054 & \textcolor{blue}{pass} / \textcolor{blue}{pass} / \textcolor{blue}{pass}  \\ 
& \cellcolor{lightgreen} LinkNet* & 0.0007 / 31.7815 & 0.0011 / 30.2647 & 0.9998 & -0.0089 & 0.0015 & 0.0245 & \textcolor{blue}{pass} / \textcolor{blue}{pass} / \textcolor{blue}{pass}  \\ 
& \cellcolor{lightgreen} FPN* & 0.005 / 23.7018 & 0.0051 / 23.6449 & 1.0000 & -0.0142 & -0.0207 & 0.0268 & \textcolor{blue}{pass} / \textcolor{blue}{pass} / \textcolor{blue}{pass}  \\ 
& \cellcolor{lightgreen} PSPNet* & 0.0099 / 20.6672 & 0.0097 / 20.7976 & 1.0000 & -0.0064 & -0.0116 & 0.0231 & \textcolor{blue}{pass} / \textcolor{blue}{pass} / \textcolor{blue}{pass}  \\ 
& \cellcolor{lightgreen} PAN* & 0.0049 / 23.8625 & 0.0049 / 23.8631 & 0.9981 & -0.0266 & 0.0044 & 0.0278 & \textcolor{blue}{pass} / \textcolor{blue}{pass} / \textcolor{blue}{pass}  \\ 
& \cellcolor{lightgreen} LRASPP* & 0.0094 / 20.9564 & 0.0096 / 20.8695 & 1.0000 & -0.0210 & -0.0148 & 0.0329 & \textcolor{blue}{pass} / \textcolor{blue}{pass} / \textcolor{blue}{pass}  \\ 
& \cellcolor{lightgreen} DeeplabV3* & 0.0093 / 20.9778 & 0.0093 / 21.0045 & 1.0000 & -0.0058 & 0.0021 & 0.0287 & \textcolor{blue}{pass} / \textcolor{blue}{pass} / \textcolor{blue}{pass}  \\ 
& \cellcolor{lightred} Baseline 1 & 0.0008 / 31.047 & 0.0003 / 36.2516 & 0.9999 & 1.0000 & 1.0000 & 1.0000 & \textcolor{blue}{pass} / \textcolor{blue}{pass} / \textcolor{red}{fail}  \\ 
& \cellcolor{lightred} Baseline 2 & 0.0008 / 31.047 & 0.0003 / 35.8167 & 1.0000 & 1.0000 & 1.0000 & 1.0000 & \textcolor{blue}{pass} / \textcolor{blue}{pass} / \textcolor{red}{fail}  \\

\bottomrule
\end{tabularx}}
\end{table*}

\paragraph{Triggers and Responses/Watermarks}
\label{sec:triggers_responses}
We consider four different methods to construct the trigger inputs $\inputwater$ and three corresponding response types $\outwater$/$\watermark$. The visualization of triggers and responses considered in this work are shown in \figref{fig:trigger_ex}. A \textbf{patch} is a single color box that is smaller than the size of the image \cite{li2022backdoor, liu2020survey}. While the only requirement for the patch trigger is that it is unique, we simplify the patch to 5 locations (top left, top right, center, bottom left, and bottom right) and 3 sizes (small, quarter, and half). Figure (a) shows a small, purple patch in the top left of the image, and figure (e) shows a quarter, orange patch in the bottom right of the image. A \textbf{block} (b, f) is a single color box that is the size of the image. \textbf{Noise} (c) is a randomly generated image of noise taken from a uniform sampling. \textbf{Steganography} (steg, d) is an image that contains secret information embedded within the image (secret/discrete). For this work, we create the steg images using the least significant bit (LSB) embedding method and an arbitrary secret image of the SatML logo, i.e. ($\inputwater = \text{LSB}(\inputclean, \text{SatML})$). An \textbf{image} (g) is simply a random image that is not contained in the training or test set, and finally, an inverse (l) is the inverse of a ground truth segmentation mask.

\paragraph{Metrics}
\label{sec:image_metrics} 
We consider different metrics and evaluation criteria depending upon the property of fragile watermarking to be evaluated. For \textbf{fidelity}, we use mean squared error (MSE) and peak-signal-to-noise ratio (PSNR) metrics for image reconstruction tasks and intersection over union (IoU) for semantic segmentation tasks. In the evaluation, fidelity is considered a \textcolor{blue}{pass} if the watermarked models achieve performance metrics that are comparable to those of the original, non-watermarked models, or rather $\text{MSE}(\model(\inputIm), \model(\out)) - \text{MSE}(\watermarkedModel(\inputIm), \watermarkedModel(\out)) < 0.02$, indicating that the embedding process has not significantly degraded the model's functionality or output quality. This value was chosen to be comparable to the NCC threshold set for retrievability and fragility. In the \textit{Results}, we simplify this evaluation notation to: ($\mid\model_{mse} - \watermarkedModel_{mse}\mid < 0.02$).

\textbf{Retrievability} and \textbf{Fragility} metrics utilize the normal cross-correlation (NCC) metric for authentication and verification described in \eqref{eq:ncc}, where $x_1$ and $x_2$ are the images being compared, $\mu$ is the mean pixel value, and $\sigma$ is the standard deviation of the pixel values. NCC is a statistical measure of how closely two images resemble each other while being invariant to changes in the brightness or intensity of an image and is a commonly used metric in watermarking tasks. NCC ranges from -1 to 1, where a value of 1 indicates that the two images are the same, a value of -1 indicates that the images are complete opposites, and a value of 0 indicates that the images are not similar.
Retrievability is considered a \textcolor{blue}{pass} if $\text{NCC}(w_f, \watermarkedModel(\inputwater)) > 0.95$, meaning that the output of the model resulting from a trigger image should resemble the watermark. Fragility, however, is considered a \textcolor{blue}{pass} when this relationship has broken, or rather $\text{NCC}(w_f, \attackedModel(\inputwater)) < 0.95$. The NCC threshold value is chosen to resemble previous image processing watermark works \cite{zhang2020model,quan2020watermarking}.%
\begin{equation}
    \label{eq:ncc}
    \text{NCC}(x_1, x_2) = \frac{(x_1 - \mu_{x_1}) \cdot (x_2 - \mu_{x_2})}{N \sigma_{x_1} \sigma_{x_2}}
\end{equation}%
%
\subsection{Baselines}
As this is the first work to consider fragile watermarking for image transformation models, we are not able to compare against other fragile watermarking methods. We do, however, compare against two robust baselines derived from~\cite{zhang2020model} to demonstrate the fragility of this proposed method compared to these methods. \\

\noindent\textbf{Baseline 1:} A Hide model $\hide$ learns to embed a watermark into a clean image to create a container image, i.e., $\hide(\watermark, \inputclean) = \container$. This container is minimally different from the clean image and contains the watermark. A Reveal model $\reveal$ then learns to extract the embedded watermark image from a container, i.e., $\reveal(\container) = \outreveal \approx \watermark$. Using a Hide model and a Reveal model, we adopt the framework from \cite{zhang2020model} and train the target model to reconstruct container outputs from container and clean image inputs. The Reveal model is then used to extract the embedded watermark from the output of the target model. If a clean image is inserted into the Reveal model, the model will respond with a blank image. The code for Baseline 1 is available here: \textbf{\url{https://github.com/ZJZAC/Deep-Model-Watermarking}}.\\

\noindent\textbf{Baseline 2:} By moving the Hide model to process outputs of the target model, we can create an additional spatial watermarking scheme for image processing tasks. For Baseline 2, the target model is trained for the desired image processing task, i.e, $\inputclean \to \outclean$. The output of the target model is then combined with the desired watermark, and the Hide model maps this image pair to a container of the processed image. The watermark is then retrievable from the Reveal model. Baseline 2 is treated as a black box system, where a user of the system is unaware of the Hide task, which adds the watermark before being returned to the user.

\subsection{Implementation Details}
To evaluate the proposed approach's ability to fragile watermark a model, we train the seven different image transformations models from \secref{sec:models} using the training approach described in \secref{sec:immediate_train} for immediate tasks (reconstruction). These models are trained with the following hyperparameters: $\text{batchsize}=32$, $\text{learning rate}=0.001$, $\text{epochs}=10$, $\text{optimizer} = \text{Adam}$. We use a patch-to-block trigger-response scheme, where the patch is a small purple box in the top left corner shown by \figref{fig:trigger_ex}a and the block is a green image block shown by \figref{fig:trigger_ex}f. We evaluate this approach against the two baselines described above using the CIFAR-10, ImageNet, and CLWD datasets for \textbf{fidelity}, \textbf{retrievability}, and \textbf{fragility}. 

In our evaluations, fidelity and retrievability are assessed prior to any alterations to the trained models. After collecting these metrics, we then apply three different black box attacks: 1) finetune for 1 epoch \textbf{(ftune1)}, 2) finetune for 5 epochs \textbf{(ftune5)}, and 3) \textbf{overwrite }the fragile watermark with a new trigger-response pair. We then evaluate these altered models for fragility.

All experiments were conducted on a macOS Monterey 12.5.1 with a 2.3 GHz 8-Core Intel Core i9 processor with 16 GB 2667 MHz DDR4 of memory.

\begin{figure*}[tb!]
    \centering
    \includegraphics[width=\textwidth]{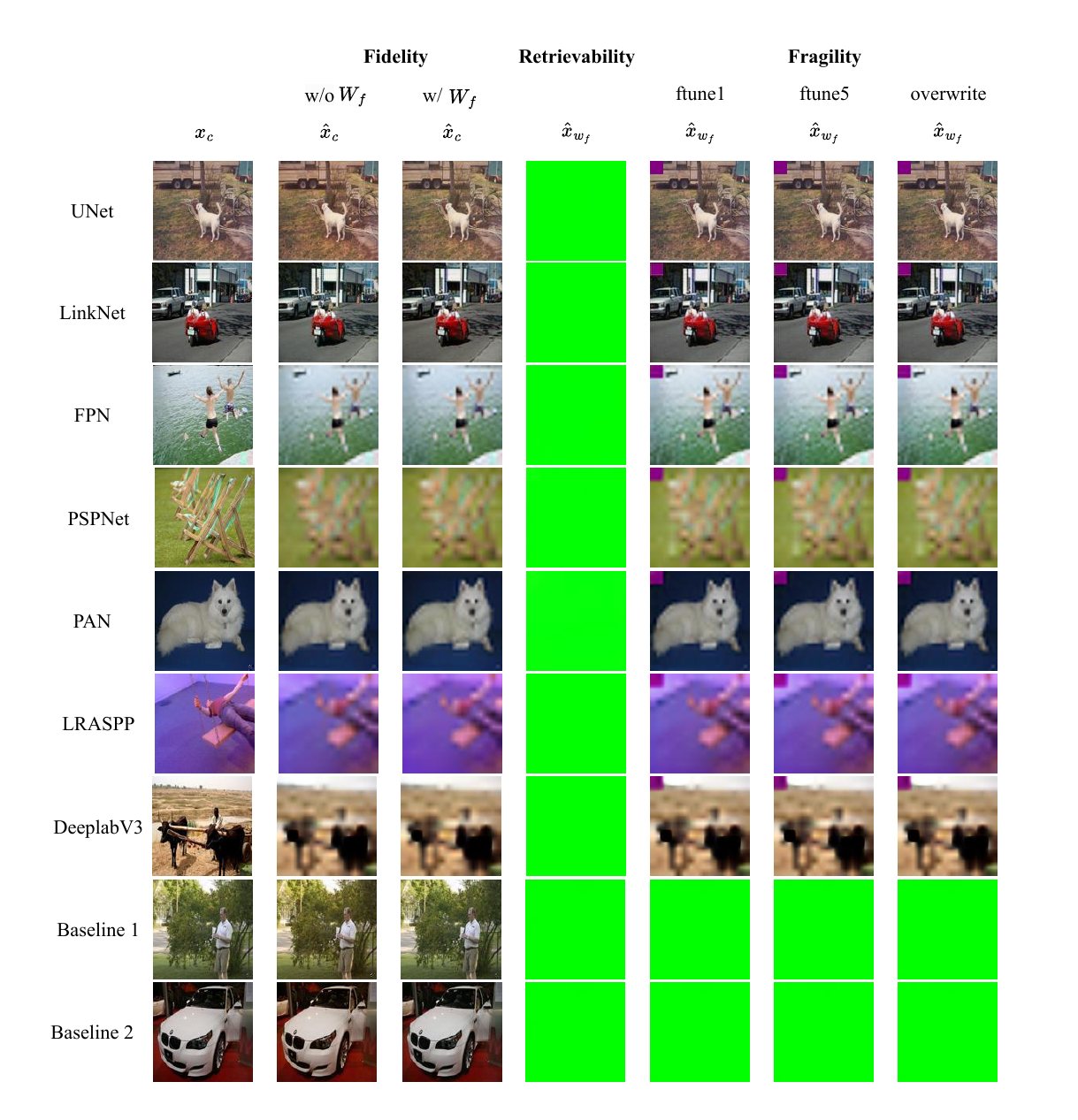}
    \caption{Fidelity, retrievability, and fragile watermarking results for models fragile watermarked with the proposed approach using a patch-to-block trigger scheme (UNet, LinkNet, FPN, PSPNet, PAN, LRASPP, and Deeplabv3) as well as two baselines. The left-most column shows a clean input $\inputclean$ which should resemble the reconstructed image from a model without the watermark (\textit{w/o $\watermarking_f$}) and the reconstructed image from a model with the watermark (\textit{w/ $\watermarking_f$}). The middle column showcases the reconstructed output from a trigger input. This image should be the green block watermark. The last three columns show the output from a trigger image after the corresponding attacks: \textit{ftune1}, \textit{ftune5}, \textit{overwrite}. All models utilizing the proposed approach meet the fidelity, retrievability, and fragility outcomes. The two baselines, however, are not fragile, as indicated by the persistent watermark after each attack.}
    \label{fig:img_results_all}
\end{figure*}

\begin{table}[tb!]
\caption{Security of the watermark in the presence of fake triggers. An \textcolor{red}{\ding{55}} indicates that the true watermark is not retrieved. This means that the trigger-watermark relationship is secure against fake triggers.}
\label{tab:rq6_defense}
\centering
\begin{tabular}{ll|cc}
\toprule
\multicolumn{2}{c}{Patch} & \multicolumn{2}{c}{Retrievability} \\

\cline{0-3}
Color    & Location       & NCC             & Eval             \\

\midrule
purple & top-right & 0.0207 & \textcolor{red}{\ding{55}} \\ 
purple & center & 0.0208 & \textcolor{red}{\ding{55}} \\ 
purple & bottom-left & 0.0205 & \textcolor{red}{\ding{55}} \\ 
purple & bottom-right & 0.0201 & \textcolor{red}{\ding{55}} \\ 
blue & top-left & 0.0196 & \textcolor{red}{\ding{55}} \\ 
green & top-left & 0.0607 & \textcolor{red}{\ding{55}} \\ 
pink & top-left & 0.0327 & \textcolor{red}{\ding{55}} \\ 
orange & top-left & 0.0413 & \textcolor{red}{\ding{55}} \\ 
yellow & top-left & 0.0490 & \textcolor{red}{\ding{55}} \\ 

\bottomrule
\end{tabular}
\end{table}
\begin{table}[tb!]
\caption{Fragile watermarking metrics and evaluation for various $\alpha$ weights, corresponding to the weight of the watermark during training.}
\label{tab:rq4}
\begin{tabularx}{\columnwidth}{
    >{\centering\arraybackslash}p{0.75cm}| 
    >{\centering\arraybackslash}p{1.15cm}| 
    >{\centering\arraybackslash}p{1.00cm} 
    >{\centering\arraybackslash}p{1.00cm} 
    >{\centering\arraybackslash}p{1.00cm}|
    >{\centering\arraybackslash}p{1.00cm}}  

\toprule

\multirow{2}{*}{\textbf{Alpha}}  
& \textbf{Fidelity}
& \multicolumn{3}{c|}{\textbf{Fragility ($\watermarking$ Authentication)}}                 
& \multirow{2}{*}{\textbf{Eval}}           \\                             
\cline{2-5}

& \textit{$\watermarking$ Auth.}
& \textit{ftune1}
& \textit{ftune5}
& \textit{overwrite }\\

& (NCC)
& (NCC)
& (NCC)
& (NCC) \\

\midrule
0.0 & 0.1209 &  -  &  -  &  -  &  \textcolor{red}{\ding{55}}\\ 
0.2 & 0.9984 & 0.0173 & 0.0245 & 0.0229 & \textcolor{green}{\ding{51}} / \textcolor{green}{\ding{51}}  \\ 
0.4 & 0.9990 & 0.0242 & 0.0215 & 0.0236 & \textcolor{green}{\ding{51}} / \textcolor{green}{\ding{51}}  \\ 
0.6 & 0.9991 & 0.0490 & 0.0218 & 0.0240 & \textcolor{green}{\ding{51}} / \textcolor{green}{\ding{51}}  \\ 
0.8 & 0.9991 & 0.0290 & 0.0249 & 0.0249 & \textcolor{green}{\ding{51}} / \textcolor{green}{\ding{51}}  \\ 
1.0 & 0.9993 & 0.0463 & 0.0220 & 0.0237 & \textcolor{green}{\ding{51}} / \textcolor{green}{\ding{51}}  \\

\bottomrule
\end{tabularx}
\end{table}

\subsection{Results}
\tabref{tab:rq1} shows the fidelity, retrievability, and fragility results for the various image transformation models and baselines. In regard to fidelity, we observe that each of the models---including the baselines---does not experience a significant decrease in performance when comparing the MSE and PSNR for reconstruction tasks between models without fragile watermarking and those with the watermarking process ($\mid\model_{mse} - \watermarkedModel_{mse}\mid < 0.02$). While some models do not achieve high reconstruction performance, this is expected because these models are designed primarily for semantic segmentation rather than image reconstruction. Our aim is not to develop high-performing reconstruction models but to test the effectiveness of the fragile watermarking approach. The minimal impact on performance metrics indicates that the fidelity of the watermarking process is successful (\textcolor{blue}{pass}).

\begin{table*}[tb!]
\caption{Fidelity, retrievability, and fragility metrics for various triggers and watermarks/responses.}
\label{tab:rq2}
\resizebox{0.95\textwidth}{!}{
    \begin{tabularx}{\textwidth}{
    p{1.25cm}| 
    >{\centering\arraybackslash}p{0.75cm}
    >{\centering\arraybackslash}p{0.65cm}| 
    >{\centering\arraybackslash}p{2.0cm}  
    >{\centering\arraybackslash}p{2.0cm} |
    >{\centering\arraybackslash}p{1.5cm}| 
    >{\centering\arraybackslash}p{1.15cm} 
    >{\centering\arraybackslash}p{1.15cm} 
    >{\centering\arraybackslash}p{1.15cm}|
    >{\centering\arraybackslash}p{2.25cm}}

\toprule
                      
\multirow{3}{*}{\centering\textbf{Model}} 
& \multirow{3}{*}{\textbf{Trigger}}
& \multirow{3}{*}{\textbf{$\watermarking$}}
& \multicolumn{2}{c|}{\textbf{Fidelity}} 
& \textbf{Retrievability}
& \multicolumn{3}{c|}{\textbf{Fragility ($\watermarking$ Authentication)}}                 
& \multirow{2}{*}{\textbf{Fragile Watermark}}           \\                             
\cline{4-9}

&
&
& \textit{Recon. w/o $\fragileWatermarking$}
& \textit{Recon. w/ $\fragileWatermarking$}
& \textit{$\watermarking$ Auth.}
& \textit{ftune1}
& \textit{ftune5}
& \textit{overwrite }\\

&
&
& (mse/psnr)
& (mse/psnr)
& (NCC)
& (NCC)
& (NCC)
& (NCC)
& (Fidelity/Retrieve/Fragility) \\

\midrule
\multirow{7}{*}{UNet} & \cellcolor{lightgreen} patch & \cellcolor{lightgreen} block & 0.0004 / 34.612 & 0.0005 / 32.9604 & 1.0000 & 0.0282 & 0.0196 & 0.0063 & \textcolor{blue}{pass} / \textcolor{blue}{pass} / \textcolor{blue}{pass}  \\ 
& \cellcolor{lightgreen} block & \cellcolor{lightgreen} block & 0.0004 / 34.612 & 0.0004 / 34.6334 & 1.0000 & -0.9557 & -0.9869 & -0.9974 & \textcolor{blue}{pass} / \textcolor{blue}{pass} / \textcolor{blue}{pass}  \\ 
& \cellcolor{lightgreen} noise & \cellcolor{lightgreen} block & 0.0004 / 34.612 & 0.0004 / 35.0595 & 1.0000 & 0.0160 & 0.0207 & -0.0213 & \textcolor{blue}{pass} / \textcolor{blue}{pass} / \textcolor{blue}{pass}  \\ 
& \cellcolor{lightgreen} steg & \cellcolor{lightgreen} block & 0.0004 / 34.612 & 0.0006 / 32.7577 & 0.9996 & 0.0463 & 0.0593 & -0.0197 & \textcolor{blue}{pass} / \textcolor{blue}{pass} / \textcolor{blue}{pass}  \\ 
& \cellcolor{lightgreen} patch & \cellcolor{lightgreen} patch & 0.0004 / 34.612 & 0.0003 / 35.3914 & 0.9970 & 0.6037 & 0.5997 & 0.6003 & \textcolor{blue}{pass} / \textcolor{blue}{pass} / \textcolor{blue}{pass}  \\ 
& \cellcolor{lightgreen} patch & \cellcolor{lightgreen} block & 0.0004 / 34.612 & 0.0005 / 32.9604 & 1.0000 & 0.0282 & 0.0196 & 0.0063 & \textcolor{blue}{pass} / \textcolor{blue}{pass} / \textcolor{blue}{pass}  \\ 
& \cellcolor{lightgreen} patch & \cellcolor{lightgreen} image & 0.0004 / 34.612 & 0.0005 / 33.5196 & 0.9993 & 0.0463 & 0.0220 & 0.0237 & \textcolor{blue}{pass} / \textcolor{blue}{pass} / \textcolor{blue}{pass}  \\ 

\midrule
\multirow{7}{*}{LinkNet} & \cellcolor{lightgreen} patch & \cellcolor{lightgreen} block & 0.0007 / 32.1809 & 0.0012 / 29.8473 & 0.9999 & 0.0484 & 0.0073 & 0.0241 & \textcolor{blue}{pass} / \textcolor{blue}{pass} / \textcolor{blue}{pass}  \\ 
& \cellcolor{lightgreen} block & \cellcolor{lightgreen} block & 0.0007 / 32.1809 & 0.0027 / 26.4462 & 1.0000 & -0.3288 & -0.9823 & -0.9686 & \textcolor{blue}{pass} / \textcolor{blue}{pass} / \textcolor{blue}{pass}  \\ 
& \cellcolor{lightgreen} noise & \cellcolor{lightgreen} block & 0.0007 / 32.1809 & 0.001 / 30.9468 & 0.9999 & -0.0181 & 0.0221 & -0.0188 & \textcolor{blue}{pass} / \textcolor{blue}{pass} / \textcolor{blue}{pass}  \\ 
& \cellcolor{lightgreen} steg & \cellcolor{lightgreen} block & 0.0007 / 32.1809 & 0.0014 / 29.4594 & 0.9992 & 0.0499 & 0.0231 & -0.0108 & \textcolor{blue}{pass} / \textcolor{blue}{pass} / \textcolor{blue}{pass}  \\ 
& \cellcolor{lightgreen} patch & \cellcolor{lightgreen} patch & 0.0007 / 32.1809 & 0.0007 / 32.4587 & 0.9940 & 0.5992 & 0.5986 & 0.5990 & \textcolor{blue}{pass} / \textcolor{blue}{pass} / \textcolor{blue}{pass}  \\ 
& \cellcolor{lightgreen} patch & \cellcolor{lightgreen} block & 0.0007 / 32.1809 & 0.0012 / 29.8473 & 0.9999 & 0.0484 & 0.0073 & 0.0241 & \textcolor{blue}{pass} / \textcolor{blue}{pass} / \textcolor{blue}{pass}  \\ 
& \cellcolor{lightgreen} patch & \cellcolor{lightgreen} image & 0.0007 / 32.1809 & 0.002 / 27.3491 & 0.9869 & 0.0229 & 0.0252 & 0.0253 & \textcolor{blue}{pass} / \textcolor{blue}{pass} / \textcolor{blue}{pass}  \\

\midrule
\multirow{7}{*}{FPN} & \cellcolor{lightgreen} patch & \cellcolor{lightgreen} block & 0.0053 / 23.6678 & 0.0054 / 23.5346 & 0.9999 & 0.0491 & 0.0190 & 0.0276 & \textcolor{blue}{pass} / \textcolor{blue}{pass} / \textcolor{blue}{pass}  \\ 
& \cellcolor{lightgreen} block & \cellcolor{lightgreen} block & 0.0053 / 23.6678 & 0.0054 / 23.5697 & 1.0000 & -0.9938 & -0.9942 & -0.9938 & \textcolor{blue}{pass} / \textcolor{blue}{pass} / \textcolor{blue}{pass}  \\ 
& \cellcolor{lightgreen} noise & \cellcolor{lightgreen} block & 0.0053 / 23.6678 & 0.0052 / 23.7491 & 1.0000 & 0.1430 & 0.0260 & -0.0152 & \textcolor{blue}{pass} / \textcolor{blue}{pass} / \textcolor{blue}{pass}  \\ 
& \cellcolor{lightgreen} steg & \cellcolor{lightgreen} block & 0.0053 / 23.6678 & 0.0055 / 23.4904 & 0.9997 & 0.0931 & 0.0278 & -0.0161 & \textcolor{blue}{pass} / \textcolor{blue}{pass} / \textcolor{blue}{pass}  \\ 
& \cellcolor{lightgreen} patch & \cellcolor{lightgreen} patch & 0.0053 / 23.6678 & 0.0053 / 23.6858 & 0.9677 & 0.5707 & 0.5712 & 0.5690 & \textcolor{blue}{pass} / \textcolor{blue}{pass} / \textcolor{blue}{pass}  \\ 
& \cellcolor{lightgreen} patch & \cellcolor{lightgreen} block & 0.0053 / 23.6678 & 0.0054 / 23.5346 & 0.9999 & 0.0491 & 0.0190 & 0.0276 & \textcolor{blue}{pass} / \textcolor{blue}{pass} / \textcolor{blue}{pass}  \\ 
& \cellcolor{lightgreen} patch & \cellcolor{lightgreen} image & 0.0053 / 23.6678 & 0.0052 / 23.7619 & 0.9696 & 0.0358 & 0.0238 & 0.0253 & \textcolor{blue}{pass} / \textcolor{blue}{pass} / \textcolor{blue}{pass}  \\ 

\midrule
\multirow{7}{*}{PSPNet} & \cellcolor{lightgreen} patch & \cellcolor{lightgreen} block & 0.0102 / 20.6874 & 0.0102 / 20.7309 & 1.0000 & 0.0218 & 0.0165 & 0.0280 & \textcolor{blue}{pass} / \textcolor{blue}{pass} / \textcolor{blue}{pass}  \\ 
& \cellcolor{lightgreen} block & \cellcolor{lightgreen} block & 0.0102 / 20.6874 & 0.0103 / 20.6694 & 1.0000 & -0.9977 & -0.9893 & -0.9957 & \textcolor{blue}{pass} / \textcolor{blue}{pass} / \textcolor{blue}{pass}  \\ 
& \cellcolor{lightgreen} noise & \cellcolor{lightgreen} block & 0.0102 / 20.6874 & 0.0101 / 20.7634 & 1.0000 & -0.3438 & 0.1455 & -0.0276 & \textcolor{blue}{pass} / \textcolor{blue}{pass} / \textcolor{blue}{pass}  \\ 
& \cellcolor{lightgreen} steg & \cellcolor{lightgreen} block & 0.0102 / 20.6874 & 0.0103 / 20.6354 & 0.9999 & 0.0014 & 0.0689 & -0.0250 & \textcolor{blue}{pass} / \textcolor{blue}{pass} / \textcolor{blue}{pass}  \\
& \cellcolor{lightgreen} patch & \cellcolor{lightgreen} patch & 0.0102 / 20.6874 & 0.0101 / 20.7841 & 0.9999 & 0.8731 & 0.8666 & 0.5348 & \textcolor{blue}{pass} / \textcolor{blue}{pass} / \textcolor{blue}{pass}  \\ 
& \cellcolor{lightgreen} patch & \cellcolor{lightgreen} block & 0.0102 / 20.6874 & 0.0102 / 20.7309 & 1.0000 & 0.0218 & 0.0165 & 0.0280 & \textcolor{blue}{pass} / \textcolor{blue}{pass} / \textcolor{blue}{pass}  \\ 
& \cellcolor{lightred} patch & \cellcolor{lightred} image & 0.0102 / 20.6874 & 0.0102 / 20.693 & 0.9261 & 0.0944 & 0.0218 & 0.0247 & \textcolor{blue}{pass} / \textcolor{red}{fail} / \textcolor{blue}{pass}  \\ 

\midrule
\multirow{7}{*}{PAN} & \cellcolor{lightgreen} patch & \cellcolor{lightgreen} block & 0.0052 / 23.8267 & 0.0053 / 23.701 & 0.9999 & 0.0223 & 0.0132 & 0.0247 & \textcolor{blue}{pass} / \textcolor{blue}{pass} / \textcolor{blue}{pass}  \\ 
& \cellcolor{lightgreen} block & \cellcolor{lightgreen} block & 0.0052 / 23.8267 & 0.0053 / 23.6807 & 1.0000 & -0.9919 & -0.9933 & -0.9966 & \textcolor{blue}{pass} / \textcolor{blue}{pass} / \textcolor{blue}{pass}  \\ 
& \cellcolor{lightgreen} noise & \cellcolor{lightgreen} block & 0.0052 / 23.8267 & 0.0052 / 23.7965 & 1.0000 & 0.5400 & -0.0351 & -0.0174 & \textcolor{blue}{pass} / \textcolor{blue}{pass} / \textcolor{blue}{pass}  \\ 
& \cellcolor{lightgreen} steg & \cellcolor{lightgreen} block & 0.0052 / 23.8267 & 0.0054 / 23.5992 & 0.9998 & 0.0831 & 0.0774 & -0.0324 & \textcolor{blue}{pass} / \textcolor{blue}{pass} / \textcolor{blue}{pass}  \\ 
& \cellcolor{lightgreen} patch & \cellcolor{lightgreen} patch & 0.0052 / 23.8267 & 0.0054 / 23.5326 & 0.9578 & 0.6010 & 0.5680 & 0.5674 & \textcolor{blue}{pass} / \textcolor{blue}{pass} / \textcolor{blue}{pass}  \\ 
& \cellcolor{lightgreen} patch & \cellcolor{lightgreen} block & 0.0052 / 23.8267 & 0.0053 / 23.701 & 0.9999 & 0.0223 & 0.0132 & 0.0247 & \textcolor{blue}{pass} / \textcolor{blue}{pass} / \textcolor{blue}{pass}  \\ 
& \cellcolor{lightred} patch & \cellcolor{lightred} image & 0.0052 / 23.8267 & 0.0055 / 23.4114 & 0.9185 & 0.0548 & 0.0243 & 0.0270 & \textcolor{blue}{pass} / \textcolor{red}{fail} / \textcolor{blue}{pass}  \\ 

\midrule
\multirow{7}{*}{LRASPP} & \cellcolor{lightgreen} patch & \cellcolor{lightgreen} block & 0.0098 / 20.9127 & 0.0099 / 20.8496 & 1.0000 & -0.5521 & 0.0116 & 0.0297 & \textcolor{blue}{pass} / \textcolor{blue}{pass} / \textcolor{blue}{pass}  \\ 
& \cellcolor{lightgreen} block & \cellcolor{lightgreen} block & 0.0098 / 20.9127 & 0.0097 / 20.9408 & 0.9999 & -0.9808 & -0.9861 & -0.9974 & \textcolor{blue}{pass} / \textcolor{blue}{pass} / \textcolor{blue}{pass}  \\ 
& \cellcolor{lightgreen} noise & \cellcolor{lightgreen} block & 0.0098 / 20.9127 & 0.0097 / 20.9515 & 0.9999 & -0.5699 & 0.0878 & -0.0166 & \textcolor{blue}{pass} / \textcolor{blue}{pass} / \textcolor{blue}{pass}  \\ 
& \cellcolor{lightgreen} steg & \cellcolor{lightgreen} block & 0.0098 / 20.9127 & 0.0105 / 20.562 & 0.9966 & 0.0539 & 0.0671 & -0.0186 & \textcolor{blue}{pass} / \textcolor{blue}{pass} / \textcolor{blue}{pass}  \\ 
& \cellcolor{lightgreen} patch & \cellcolor{lightgreen} patch & 0.0098 / 20.9127 & 0.0099 / 20.8446 & 0.9984 & 0.8710 & 0.8647 & 0.5394 & \textcolor{blue}{pass} / \textcolor{blue}{pass} / \textcolor{blue}{pass}  \\ 
& \cellcolor{lightgreen} patch & \cellcolor{lightgreen} block & 0.0098 / 20.9127 & 0.0099 / 20.8496 & 1.0000 & -0.5521 & 0.0116 & 0.0297 & \textcolor{blue}{pass} / \textcolor{blue}{pass} / \textcolor{blue}{pass}  \\ 
& \cellcolor{lightred} patch & \cellcolor{lightred} image & 0.0098 / 20.9127 & 0.0096 / 20.9966 & 0.8986 & 0.1673 & 0.1926 & 0.0217 & \textcolor{blue}{pass} / \textcolor{red}{fail} / \textcolor{blue}{pass}  \\ 

\midrule
\multirow{7}{*}{DeeplabV3} & \cellcolor{lightgreen} patch & \cellcolor{lightgreen} block & 0.0095 / 21.0361 & 0.0098 / 20.9003 & 1.0000 & 0.0289 & 0.0227 & 0.0271 & \textcolor{blue}{pass} / \textcolor{blue}{pass} / \textcolor{blue}{pass}  \\ 
& \cellcolor{lightgreen} block & \cellcolor{lightgreen} block & 0.0095 / 21.0361 & 0.0096 / 20.9885 & 1.0000 & -0.9988 & -0.9972 & -0.9980 & \textcolor{blue}{pass} / \textcolor{blue}{pass} / \textcolor{blue}{pass}  \\ 
& \cellcolor{lightgreen} noise & \cellcolor{lightgreen} block & 0.0095 / 21.0361 & 0.0096 / 21.0052 & 1.0000 & 0.1767 & 0.1019 & -0.0230 & \textcolor{blue}{pass} / \textcolor{blue}{pass} / \textcolor{blue}{pass}  \\ 
& \cellcolor{lightgreen} steg & \cellcolor{lightgreen} block & 0.0095 / 21.0361 & 0.012 / 20.1584 & 0.9963 & 0.0696 & 0.0585 & -0.0193 & \textcolor{blue}{pass} / \textcolor{blue}{pass} / \textcolor{blue}{pass}  \\ 
& \cellcolor{lightgreen} patch & \cellcolor{lightgreen} patch & 0.0095 / 21.0361 & 0.0096 / 20.9631 & 1.0000 & 0.8716 & 0.8768 & 0.5432 & \textcolor{blue}{pass} / \textcolor{blue}{pass} / \textcolor{blue}{pass}  \\ 
& \cellcolor{lightgreen} patch & \cellcolor{lightgreen} block & 0.0095 / 21.0361 & 0.0098 / 20.9003 & 1.0000 & 0.0289 & 0.0227 & 0.0271 & \textcolor{blue}{pass} / \textcolor{blue}{pass} / \textcolor{blue}{pass}  \\ 
& \cellcolor{lightred} patch & \cellcolor{lightred} image & 0.0095 / 21.0361 & 0.0096 / 20.9847 & 0.9382 & 0.0350 & 0.0267 & 0.0271 & \textcolor{blue}{pass} / \textcolor{red}{fail} / \textcolor{blue}{pass}  \\ 

\bottomrule

\end{tabularx}}
\end{table*}

Prior to an attack, the retrievability of the watermark is also a \textcolor{blue}{pass} as the average NCC of each model is $> 0.95$. After the attacks, however, each of the models using the proposed approach (indicated by *), are unable to retrieve the watermark as the $\text{NCC} < 0.95$. This is not the case for the baselines. For each of the baselines, the attacks do not affect the watermark retrieval. This is expected, though, as baseline 1 and baseline 2 are robust watermarking methods. This result helps to highlight the difference between robust and fragile watermarking methods.

The image results for this experiment are shown in \figref{fig:img_results_all}. Whereas the watermark of a green block is retrievable at test time prior to an attack, for all models using the proposed approach (UNet through DeeplabV3), the watermark is not retrieved after an attack. This is not the case, however, for the two baselines, as highlighted by the green blocks in the last the three image columns.

As the proposed approach achieves the fidelity, retrievability, and fragility criteria across all datasets and models for immediate tasks, we consider it a successful fragile watermarking scheme that can detect unauthorized modifications to the models while maintaining their original performance prior to any attacks.

\section{Analysis of Method Capabilities}
In this section, we present ablation experiments to further analyze the proposed approach.

\subsection{Security of the Trigger Activation}
\label{sec:security}
To evaluate the robustness of our watermarking scheme against unauthorized activations, we test the model with various fake triggers differing in color and location. \tabref{tab:rq6_defense} presents the results of this experiments. The fake triggers included patches of different colors (purple, blue, green, pink, orange, and yellow) placed at various locations on the image such as top-right, center, bottom-left, bottom-right, and top-left.

As shown in the table, all fake triggers resulted in low NCC values ranging from 0.0196 to 0.0607. The evaluation column (\textit{Eval}) consistently shows a red cross (\textcolor{red}{\ding{55}}), indicating that the watermark was not retrieved in any of these cases. These results confirm that the watermark is secure against unauthorized triggers. Only the specifically trained trigger input of a purple box in the top-left corner successfully activates the watermark, ensuring that fake triggers cannot compromise the integrity of the watermarking scheme.

\subsection{Flexibility of the Trigger}
In the previous section, we utilize a patch trigger of a purple box in the top left corner. To determine the flexibility of the trigger for this fragile watermarking process, we compare \textless trigger\textgreater{}-block performance on the ImageNet dataset, where \textless trigger\textgreater{} is either a patch, block, noise, or steganography (see \secref{sec:triggers_responses}). 

The fidelity, retrievability, and fragility results for these experiments are shown in \tabref{tab:rq2}. Prior to an attack, each of the proposed trigger types maintains performance when compared to its non-watermarking counterpart, as indicated by the MSE and PSNR metrics in the \textit{Fidelity} columns. Fidelity is, therefore, a \textcolor{blue}{pass} across each trigger type as ($\mid\model_{mse} - \watermarkedModel_{mse}\mid < 0.02$). While each trigger type performs well, the noise triggers cause the least disturbance to the target task. 

Regarding retrievability, each of the proposed triggers achieves the authentication criteria ($\text{NCC} > 0.95$). After an attack, the models all achieve the fragility criteria as well. This means that the proposed fragile watermarking scheme is flexible in regard to the trigger type, and the previous results in \tabref{tab:rq1} are not coupled to the patch trigger type.

\subsection{Flexibility of the Response/Watermark}
To decouple the response/watermark used in the fragile watermarking process, we repeat the experiment above with different types of watermarks.  We compare patch-\textless watermark\textgreater{} performance on the ImageNet dataset, where \textless watermark\textgreater{} is either a patch, block, or image (see \secref{sec:triggers_responses}). The image used for the `image' watermark is the flower shown in \figref{fig:trigger_ex}(g). 

\begin{figure*}[tb!]
    \centering
    \includegraphics[width=\textwidth]{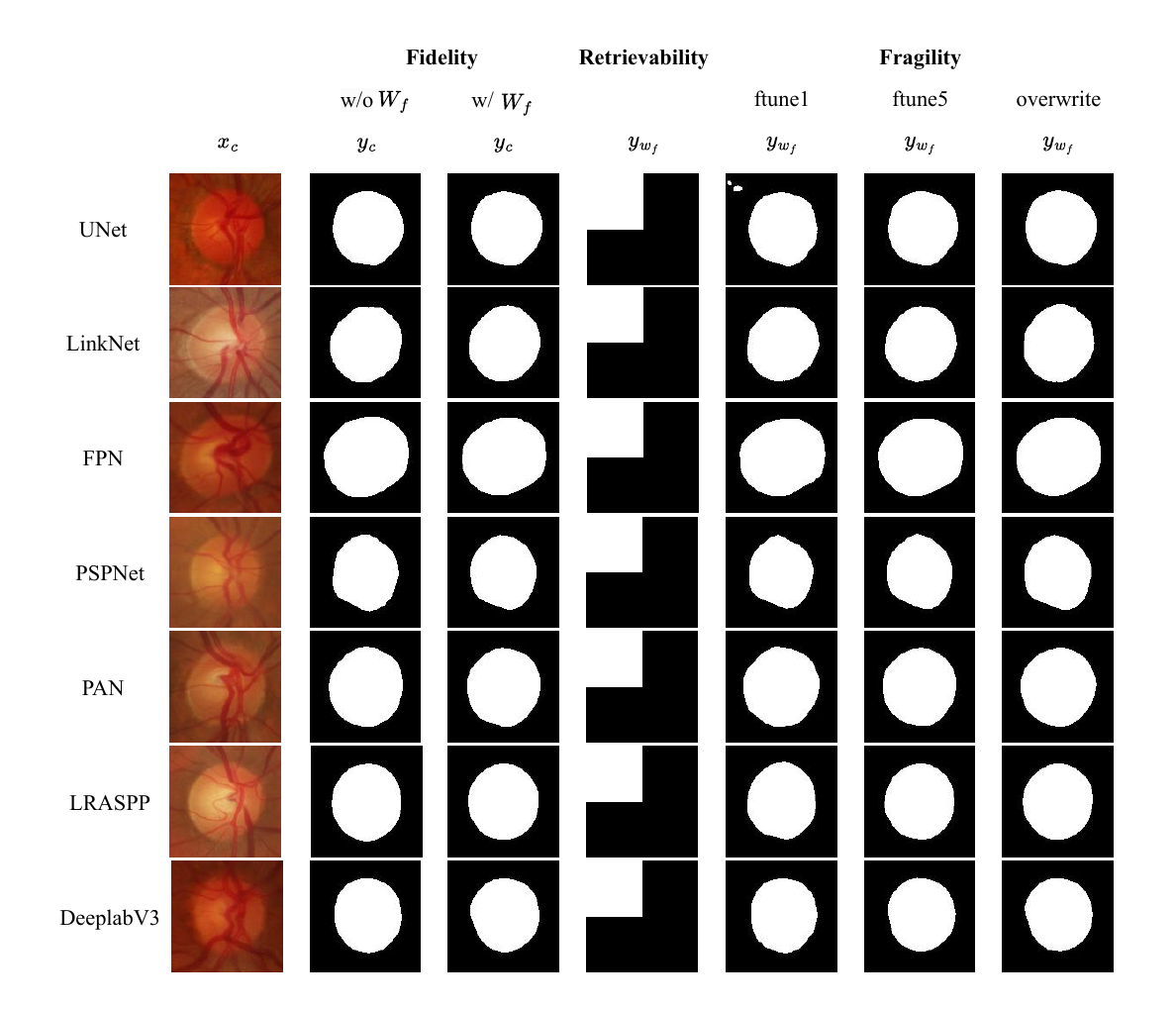}
    \caption{The image results for our approach applied to a downstream semantic segmentation task. Before an attack, clean inputs perform well on the target segmentation task, and trigger inputs $\inputwater$ result in the intended patch watermark, as shown by the middle column. After each attack, the watermark is no longer retrievable by the trigger, as shown by the last three columns.}
    \label{fig:seg_img_result}
\end{figure*}

\begin{table*}[tb!]
\caption{Results for the proposed fragile watermark approach on a downstream semantic segmentation task.}
\label{tab:rq5}
\begin{tabularx}{\textwidth}{
    p{1.25cm}| 
    >{\centering\arraybackslash}X 
    >{\centering\arraybackslash}X |
    >{\centering\arraybackslash}p{1.5cm}| 
    >{\centering\arraybackslash}p{1.15cm} 
    >{\centering\arraybackslash}p{1.15cm} 
    >{\centering\arraybackslash}p{1.15cm}|
    >{\centering\arraybackslash}X}

\toprule
\multirow{3}{*}{\centering\textbf{Model}} 
& \multicolumn{2}{c|}{\textbf{Fidelity}} 
& \textbf{Retrievability}
& \multicolumn{3}{c|}{\textbf{Fragility ($\watermarking$ Authentication)}}                 
& \multirow{2}{*}{\textbf{Fragile Watermark}}           \\                             
\cline{2-7}

& \textit{Segmentation w/o $\fragileWatermarking$}
& \textit{Segmentation w/ $\fragileWatermarking$}
& \textit{$\watermarking$ Auth.}
& \textit{ftune1}
& \textit{ftune5}
& \textit{overwrite }\\

& (IoU)
& (IoU)
& (NCC)
& (NCC)
& (NCC)
& (NCC)
& (Fidelity/Retrieve/Fragility) \\
\midrule 
\cellcolor{lightgreen} UNet & 0.9212 & 0.9212 & 1.0000 & 0.1010 & 0.0097 & 0.0174 & \textcolor{blue}{pass} / \textcolor{blue}{pass} \ \textcolor{blue}{pass}  \\ 
\cellcolor{lightgreen} LinkNet & 0.9229 & 0.9229 & 0.9999 & 0.4518 & 0.4761 & 0.2974 & \textcolor{blue}{pass} / \textcolor{blue}{pass} \ \textcolor{blue}{pass}\\ 
\cellcolor{lightgreen} FPN & 0.9227 & 0.9227 & 1.0000 & 0.0625 & 0.0210 & 0.1289 & \textcolor{blue}{pass} / \textcolor{blue}{pass} \ \textcolor{blue}{pass}  \\ 
\cellcolor{lightgreen} PSPNet & 0.9164 & 0.9164 & 0.9884 & 0.7254 & 0.6423 & 0.3367 & \textcolor{blue}{pass} / \textcolor{blue}{pass} \ \textcolor{blue}{pass}  \\ 
\cellcolor{lightgreen} PAN & 0.9212 & 0.9212 & 0.9992 & 0.0358 & 0.0439 & 0.0565 & \textcolor{blue}{pass} / \textcolor{blue}{pass} \ \textcolor{blue}{pass}  \\ 
\cellcolor{lightgreen} LRASPP & 0.9197 & 0.9197 & 0.9994 & 0.1163 & 0.0566 & -0.1086 & \textcolor{blue}{pass} / \textcolor{blue}{pass} \ \textcolor{blue}{pass}  \\ 
\cellcolor{lightgreen} DeeplabV3 & 0.9196 & 0.9196 & 0.9996 & 0.0723 & 0.0279 & -0.0126 & \textcolor{blue}{pass} / \textcolor{blue}{pass} \ \textcolor{blue}{pass}  \\  
\bottomrule
\end{tabularx}
\end{table*}

The fidelity, retrievability, and fragility results for this experiment are shown in \tabref{tab:rq2}. From these results, the proposed approach is watermark agnostic as well. While some of the image watermarks receive a fail for retrievability as shown by the sections highlighted in red, this is due to the limitations of the model rather than the proposed approach. The architectures of PSPNet, PAN, LRASPP, and DeeplabV3 include choices such as deep encoders, pooling operations, and classification-focused decoders, which lead to a loss of the fine-grained spatial and texture information necessary for high-quality image reconstruction resulting in \textcolor{red}{fails} for retrievability. The patch and block watermarks, however, are successful for all models. As such, we consider the proposed approach to be flexible to the trained watermark as well. 

\subsection{Watermarking Loss}
In this experiment, we investigate how varying the embedding strength parameter $\alpha$ affects the performance of our fragile watermarking approach. The parameter $\alpha$ controls the balance between the original model parameters and the embedded watermark during the watermarking process. A higher $\alpha$ value implies a stronger emphasis on the watermark, potentially affecting the model's fidelity, while a lower $\alpha$ might result in a watermark that is too weak to detect or too robust against alterations.

Table \tabref{tab:rq4} summarizes the results of this experiments for different values of $\alpha$ ranging from 0.0 to 1.0 in increments of 0.2. The results confirm that our fragile watermarking approach is effective when the embedding strength $\alpha$ is greater than zero. A minimal embedding strength ($\alpha = 0.2$) is sufficient to achieve high retrievability and ensure the watermark is fragile against common model modifications. Increasing $\alpha$ beyond 0.2 does not significantly impact the retrievability or fragility, suggesting that the method is robust across a range of embedding strengths.

\subsection{Downstream Tasks}
In the previous experiments, we conduct evaluations on immediate tasks. To demonstrate the feasibility of this approach for downstream tasks, we evaluate this fragile watermarking approach on a semantic segmentation task using the RIM-ONE for deep learning dataset (RIM-ONE DL). We train each of the image transformation models on a binary semantic segmentation task, where a pixel class 0 indicates the background and a pixel class 1 indicates the disc of the eye. We implement the downstream tasks training process described in \secref{sec:downstream_train}. After initial training, we collect fidelity (IoU) and retrievability metrics. After the initial training, we attack the model by finetuning for 1 epoch (\textbf{ftune1}), finetuning for 5 epochs (\textbf{ftune5}), and overwriting the watermark with a different watermark (\textbf{overwrite}). For each attacked model, we then collect fragility metrics. 

\tabref{tab:rq5} shows the fragility, retrievability, and fidelity of the proposed approach for semantic segmentation before and after the applied perturbation. Before an attack, the applied watermarking approach performs well on the semantic segmentation and watermarking tasks for all models, as shown by the high IoU and NCC value scores. As the addition of the watermarking task does not degrade performance and the NCC of the retrieved watermarks is $> 0.95$, this approach is a \textcolor{blue}{pass} for fidelity and retrievability.  After a perturbation, the watermark successfully breaks for each attack, as shown by the low NCC values for ftune1, ftune5, and overwrite attacks in regard to \textit{Fragility}. Image results for this experiment are shown in \figref{fig:seg_img_result}. Before an attack (left-most image blocks), the model performs well on the semantic segmentation and watermarking tasks, as shown by the predicted masks and watermark output for $\inputclean$ and $\inputwater$. After an attack, however, the semantic segmentation performance remains relatively the same while breaking the watermarking output, meaning the watermark is no longer retrievable. This is indicated by three right-most columns of the figure for \textit{ftune1}, \textit{ftune5}, and \textit{overwrite}. From these results, we consider the proposed fragile watermarking scheme a successful fragile approach for downstream tasks that can detect unauthorized modifications to the models while maintaining their original performance prior to any attacks.

\section{Limitations}
One limitation of the proposed fragile model watermarking scheme is its vulnerability to informed adversaries. If an attacker is aware of the specific trigger-watermark mechanism in place, they can potentially bypass it by making unauthorized changes to the content and then reapplying the watermark check. This circumvents the protection intended by the scheme, as the adversary can modify the content while still passing the verification process. While the experiment in \secref{sec:security} helps to alleviate some of this concern, it is still a possibility. 

Another important consideration is the capability of the model to be fragily watermarked. For example, models like PSPNet, PAN, LRASPP, and DeeplabV3 did not perform well with the patch-to-image watermarking scheme, as shown in \tabref{tab:rq2}. These models involve operations that cause a loss of fine-grained features, making it difficult for an image-based watermark to succeed. Since they cannot preserve the fine details required for accurate watermark verification, they inevitably fail to pass the strict verification tests. It is important, therefore, to consider the capabilities of the model to be watermarked when choosing the expected response.

\section{Conclusion}
In this work, we introduce a novel fragile model watermarking scheme for the integrity verification of image transformation/generation models. Through various experiments, we demonstrate that our approach is fragile (fidelity, receivability, and fragile), outperforms two robust baselines, secure against fake triggers, flexible to various triggers and watermarks, resilient to even small watermarking weights, and applicable to downstream tasks.  As the importance of model integrity continues to increase, this fragile model watermarking scheme provides a crucial step forward in protecting machine learning user trust and security. We look forward to exploring this work for additional tasks, including natural language processing and object detection.

\bibliographystyle{splncs04}
\bibliography{references}

\end{document}